\documentclass[aps,prl,twocolumn,amsmath,amssymb,nofootinbib,showpacs,superscriptaddress]{revtex4-2}
\usepackage[english]{babel}
\usepackage{latexsym}
\usepackage{graphics}
\usepackage{multirow}
\usepackage{graphicx}
\usepackage{epsfig}
\usepackage{color}
\usepackage{bm}
\usepackage{hyperref}
\usepackage{slashed}
\usepackage{amsmath}
\usepackage{amssymb}
\usepackage{amsthm}
\usepackage{dcolumn}
\usepackage{algorithm}
\usepackage{algpseudocode}
\usepackage{bm}
\usepackage{braket}
\usepackage{stmaryrd}
\usepackage{booktabs}
\usepackage{hyperref}
\usepackage{color}
\usepackage{epstopdf}
\usepackage{cleveref}
\usepackage{ulem}
\usepackage[svgnames]{xcolor}
\usepackage{blindtext}
\usepackage{comment}
\hypersetup{hidelinks,colorlinks=true,allcolors=DarkBlue}

\theoremstyle{remark}

\begin{document}

\preprint{APS/123-QED}

\title{Fractal states of the Schwinger model}

\author{E.V. Petrova}
\affiliation{Russian Quantum Center, Skolkovo, Moscow 143025, Russia}
\affiliation{Institute of Science and Technology Austria, Am Campus 1, 3400 Klosterneuburg, Austria}

\author{E.S. Tiunov}
\affiliation{Russian Quantum Center, Skolkovo, Moscow 143025, Russia}
\affiliation{Quantum Research Centre, Technology Innovation Institute, Abu Dhabi, UAE}

\author{M.C. Ba\~nuls}
\affiliation{Max-Planck-Institut f\"ur Quantenoptik, Hans-Kopfermann-Str.1, 85748 Garching, Germany}
\affiliation{Munich Center for Quantum Science and Technology (MCQST), Schellingstr. 4, D-80799 Mu\"nchen}

\author{A.K. Fedorov}
\affiliation{Russian Quantum Center, Skolkovo, Moscow 143025, Russia}
\affiliation{National University of Science and Technology ``MISIS”, Moscow, Russia}

\date{\today}
\begin{abstract}
The lattice Schwinger model (SM), the discrete version of QED in 1+1 dimensions, is a well-studied test bench for lattice gauge theories.
Here we study the fractal properties of the SM.
We reveal the self-similarity of the ground state, which allows one to develop a recurrent procedure for finding the ground-state wave functions and predicting ground-state energies.
We provide the results of recurrently calculating ground-state wave functions using the fractal ansatz and automized software package for fractal image processing. 
In certain parameter regimes, just a few terms are enough for our recurrent procedure to predict ground state energies close to the exact ones for several hundreds of sites.
Our findings pave the way to understanding the complexity of calculating many-body wave functions in terms of their fractal properties
as well as finding new links between condensed matter and high-energy lattice models.
\end{abstract}

\maketitle

{\it Introduction}. 
Starting from its conception in the 1950s, the Schwinger model (SM)~\cite{Schwinger1951}, or quantum electrodynamics in one spatial dimension, has become a subject of intensive research.  
One of the reasons is that, despite being one of the simplest gauge theories, this model shares a number of non-trivial features with more complex theories, like quantum chromodynamics.
The SM is analytically solvable in the special cases of massless or non-interacting particles~\cite{Schwinger1962, Crewther1980}.
However, in the general case no exact solution is known, and the model offers a challenging test bench for both analytical and numerical techniques.
Within this context, it is quite natural that methods that were originally developed for complex condensed matter systems, 
such as tensor networks~\cite{Schollwock2011,Cirac2008,Orus2014}, and quantum technologies (for recent reviews, see e.g. Refs.~\cite{Zohar2022,Banuls2020,Banuls2020-2}),
are now extended its realm to high-energy problems.
In particular, the SM has been object of systematic tensor network simulations~\cite{Banuls2013,Orus2017,Butt2020}.
Capabilities of quantum technologies are also under active research~\cite{Zohar2022,Banuls2020,Banuls2020-2} starting from the pioneering experiment~\cite{Blatt2016}.

Identifying a specific structure in the state of quantum many-body systems can provide valuable insight to reduce the complexity of their description. 
A particular example is the self-similar structure appearing in the wave function for certain quantum many-body problems.
The qubism framework~\cite{Lewenstein2012} has proposed a systematic way for visualizing many-body wave functions by mapping them onto two-dimensional (2D) plots,
which has demonstrated that Heisenberg or the Ising model in a transverse field can be characterized as fractals~\cite{Lewenstein2012}. 
The recursive (self-similar) nature of the plot is displayed by the fact that increasing the number of particles in the considered system reflects in an increase in pattern repetitions.
For applications of fractals in studies of many-body models, see also Refs.~\cite{Olemsko1993,Nakayama2009,Matsueda2016}.

The concept of self-similarity is deeply connected to multifractality and fractal dimensions in quantum many-body physics.
One of the well-studied manifestations is the multifractality of electronic wave functions in the context of the Anderson transition~\cite{Anderson1958,Evers2008}, 
where the wave functions cannot be characterized by a single fractal dimension.
Instead, an infinite number of fractal dimensions, which is known as the multifractal spectrum, is needed~\cite{Evers2008,Janssen1998,Mirlin2000,Fyodorov2006}.
This concept can be further studied as a generic property of quantum many-body systems~\cite{Wegner1981,Altshuler1986,Monthus2016,Abanin2017,Laflorencie2019,Warzel2021} as well as 
a feature of random matrix models~\cite{Fyodorov1996,Telcs1999,Fyodorov2009,Giraud2012,Kravtsov2015,Monthus2017,Haque2019,Khaymovich2020-2,Biroli2021}.
Multifractality has been studied, first, for a large class of models~\cite{Bogomolny2012,Bogomolny2014,Luitz2014,Luitz2014-2,Shlyapnikov2016,Shlyapnikov2019,Lindinger2019,Khaymovich2019-2,Khaymovich2020,Khaymovich2021}.
Gaining insight into the structure and the complexity of quantum many-body states can be crucial to understand the possibilities and limitations regarding their analytical or numerical description.

\begin{figure*}[t!]
	\centering
	\includegraphics[width=1\linewidth]{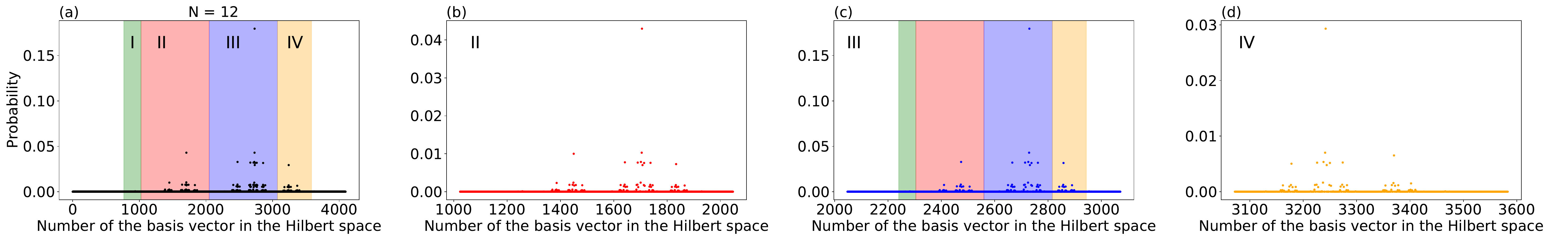}
	\vskip-1mm
	\caption{Fractal structure of the GS wave function. Here $s$ is the qubit configuration, which enumerates the basis vectors in the Hilbert space.
	In (a) the probability distribution as the function of the qubit configuration for $N=12$ sites with four highlighted regions 
	green (I), red (II), blue (III), and orange (IV) regions of the distribution is illustrated. Such regions are defined by fixing the first spins in the following way: $\ket{0011}$ -- (I), $\ket{01}$ -- (II), $\ket{10}$ -- (III), $\ket{110}$ -- (IV);
	In (b)--(d) separate plots for (II), (III), and (IV) regions, respectively, are shown.}
	\label{fig:first}
\end{figure*}

Here we explore the previously unobserved fractal structure in the ground state (GS) of the SM. 
Specifically, we show the recursive nature of the ground state wave function in the strong coupling basis, which can be clearly appreciated at the level of the corresponding probability distribution, and visualized via qubism. 
We propose a fractal ansatz and use it to construct the wave function for a larger system size starting from the exact representation for a smaller system.
Here, we make use of this feature for calculating GS wave functions and their energies.  

{\it The model}.
The SM describes 1+1 quantum electrodynamics in one time and one spatial dimension.
The discrete version of the Hamiltonian representation of the model, in the physical subspace satisfying Gauss law, can be mapped to a spin chain with long-range interactions~\cite{Susskind1976}.
For a system with open boundary conditions, the gauge degrees of freedom can be completely eliminated using Gauss law and a gauge transformation, 
yielding the following Hamiltonian:
\begin{equation}
\begin{split}
		H = &x\sum_{n = 0}^{N-2}+(\sigma^{+}_n\sigma^{-}_{n+1} + \sigma^{+}_{n+1}\sigma^{-}_{n}) \nonumber
\end{split}
\end{equation}
\begin{equation}
\begin{split}
		&+\frac{\mu}{2}\sum_{n = 0}^{N-1}(1 + (-1)^n\sigma^z_n) \\
		&+\sum_{n = 0}^{N-2}\left(\epsilon_0 + \frac{1}{2}\sum_{l=0}^n(\sigma_l^z + (-1)^l)\right)^2 = \\
		&= xV + H_{\rm diag},
	\label{eq:H1}
\end{split}
\end{equation}
where $\sigma_n^{x,y,z}$ are Pauli matrices at the site $n$ and $\sigma^{\pm} = ({\sigma^x \pm i\sigma^y})/{2}$.
The Hamiltonian is parametrized by $x ={1}/{g^2a^2}$ and $\mu = {2m}/{(g^2a)}$, where $a$ is the lattice step, $g$ is the gauge coupling, and $\epsilon_0$ is the external field, which we hereafter set to $0$.
Here $V$ is the hopping term and $H_{\rm diag}$ is the diagonal part that includes nontrivial (non-local) interaction.
In such a form, the spin up on an even site corresponds to the presence of a  fermion, and vice versa, spin down on an odd site corresponds to the presence of an antifermion. 

The SM in the spin representation preserves the global symmetries. 
It conserves the global charge symmetry, which in this notation implies $[\hat{H}, \hat{\sigma}^z_{tot}] = 0$, 
and is a consequence of the $U(1)$  gauge symmetry. 
Eigenstates have thus well defined charge, namely $\hat{\sigma}^z_{\mathrm{tot}}\ket{\psi}=s_z \ket{\psi}$, where $\hat{\sigma}^z_{\mathrm{tot}} = \sum_j \hat{\sigma}^z_{j}$.
For an even number of sites, the GS lies in the zero charge sector, or $s_z=0$, 
and we will restrict the analysis to only this sector.
This reduces the number of independent parameters to ${N \choose N/2}$.
For an odd number of sites, we explore only the sector $s_z=1/2$.

{\it Self-similar structure of the ground state}.
We demonstrate the self-similarity of the ground state wave function by representing the state $\ket{\Psi_N}$ as a vector of size $2^N$, where $N$ is the number of sites, in the computational basis. 
Each element of this vector can be chosen to be real, as far as the eigenstate is nondegenerate. 
Fig.~\ref{fig:first}(a) illustrates the corresponding probability distribution for $N = 12$ qubits with parameters $\mu = 0.1$ and $x = 1$. 

Empirically, we reveal four  GS parts that are similar to the rescaled distributions for a smaller number of sites. 
Fig.~\ref{fig:first}(b)--(d) show three out of four regions. 
Based on these observations such a self-similar structure can be used as the basis to build an ansatz.

The most compact version of it is as follows:
\begin{equation}
	\begin{gathered}
        \ket{\Psi_N^A} = W_{0011}^N\ket{0011}\ket{ \Psi_{N-4}} + W_{01}^N\ket{01}\ket{\Psi_{N-2}} +\\
        +W_{10}^N\ket{10}\ket{\Psi_{N-2}} + W_{110}^N\ket{110}\hat{T}\ket{\Psi_{N-3}} 
    \end{gathered}
    \label{eq:ansatz_4part}
\end{equation}
where $\hat{T} = \otimes_k \hat{\sigma}^x_k$ is the bit flipping operator.
$\ket{\Psi_{N-K}}$ are the GS for the $N-K$ sites, $s$ are the combination of the first $K$ spins that fix the subspace in the Hilbert space.
The combinations of the GS for smaller system size and bit strings $s$ form the basis, while  weight coefficients $W_{s}^N$ define the state in that basis.
To preserve the zero total charge for an even number of sites in Eq.~\eqref{eq:ansatz_4part}, 
the GS for the odd number has to lie in the $+1/2$ sector ($1$ is spin up and $0$ is spin down). 

The compact ansatz~\eqref{eq:ansatz_4part} presents only a first level of approximation. 
To obtain a more accurate approximation, we need to increase the number of terms in the ansatz, by taking into account more parts of the distribution.
We provide more details about advanced ansatzes in Supplementary Information. 

The fractal ansatz better approximates the true quantum state if weights do not change with the system size. 
We indeed observe that by increasing the system size weight coefficients converge (see Fig.~[4] in Supplementary Information).
Then, the structure of the GS tends to achieve the structure of the fractal meaning that  the self-similarity is independent of the system size.
Due to the fractal structure, the approximation of GS can be constructed recursively using Eq.~\eqref{eq:ansatz_4part}, (for higher order ansatzes see Supplementary Information). 
An alternative way to build GS is by learning the fractal properties of the state using 2D visualization~\cite{Lewenstein2012} and then applying a compression-decompression algorithm~\cite{Agarwal2019,pvigier_fractal} 
(for more details see Qubims framework, see Supplementary Information).

\begin{figure}[h]
	\centering
	\includegraphics[width=0.7\linewidth]{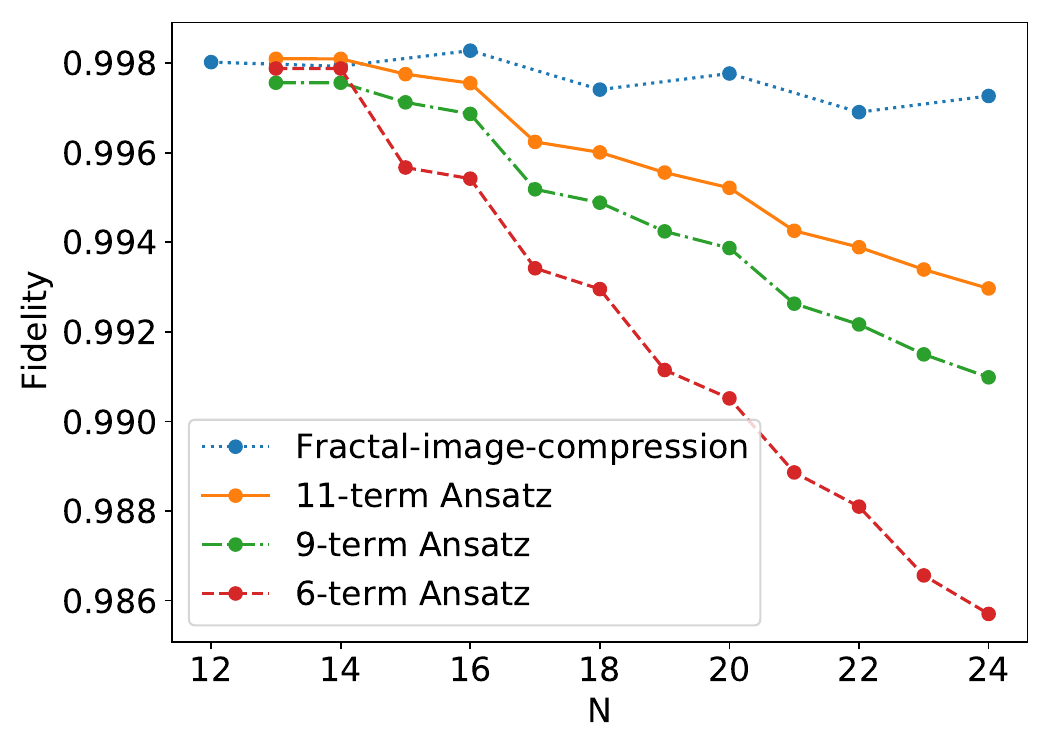}
	\vskip -3mm
	\caption{The fidelity, $\mathcal{F}=|\braket{\Psi_{\rm exact}|\Psi}|^2$, where $\Psi_{\rm exact}$ is the exact GS obtained by exact diagonalization and $\Psi$ is our prediction, is shown as a function of the system size (N) for parameters $x=1, \mu = 0.1$. 
	For the fractal-image compression algorithm, we use qubism representation of the wave function and apply the compression-decompression algorithm, decompressing it into the picture of the interested size. 
	For the recursion algorithm, we use 6, 9, and 11-term ansatzes which include 9, 11 terms respectively, and depend on the distributions for $N-2$, $N-3$, $N-4$, $N-5$, $N-6$, $N-7$, $N-8$ qubits.
	The recurrent procedure for obtaining wave functions is approximate, so the accumulation of errors leads to a decrease in fidelity.}
	\label{fig:fidelity}
\end{figure}

To demonstrate the techniques, we plot the fidelity of the reconstructed ground state as a function of the system size in Fig.~\ref{fig:fidelity}.
The starting point is chosen to be the GS for 12 qubits obtained by exact diagonalization. 
Solid, dashed-dotted, and dashed lines depict fractal ansatzes with 11, 9, and 6 terms, respectively, while the dotted curve shows the results for the image compression algorithm. 
The 11-term ansatz shows superior results, as it is a more accurate approximation compared to the 9 and 6-term ansatzes.
At the same time, we observe that both fractal methods give a bigger drop in the fidelity with the system size than the computer graphics algorithm. 
This algorithm splits the state into a larger number of parts. 
We assume that this fact leads to higher fidelity than fractal ansatzes.
Another important problem that is worth to be discussed is the error of the fractal method. 
This comes from two sources. 
First, different fractal ansatzes cover different parts of the GS distribution. 
Therefore, there will be an error if the ansatz does not cover a sector where the GS has non-vanishing weight.
The second source is how well the fractal structure of the ansatz matches that of the real state. 
The error related to the coverage can be quantified by summing squared values of weights, which should be one for a perfect ansatz (see Supplementary Information, Fig~[5]). 

Nevertheless, the two latter provide a better approximation for large $N$ (higher fidelity on Fig.~\ref{fig:fidelity}), since their structures are better fitted to the ideal GS.
A precise analysis of this second source of errors, which is the dominant one  in this case, requires additional numerical procedures, whose development is beyond the scope of the present work.

We would like to note that the same procedure can be realized for distributions associated with different values of the parameters of the model. 
The explored fractal property is also preserved in different bases (see Supplementary Information, {Fig.~[6].}) 
We also note that the fractal structure of the wave function also essentially changes across the critical point (see Supplementary Information).

{\it Ground-state energy}.
The fractal properties of the GS follow from the structure of the Hamiltonian.
Due to the translational invariance, the first two terms in the SM Hamiltonian~\eqref{eq:H1} can be rewritten as the sum of the corresponding term in the Hamiltonian for a smaller system
size plus an operator which acts only on the first few spins. 
Whereas the last diagonal term in the SM Hamiltonian~\eqref{eq:H1} does not exhibit the same symmetry, it can be split in a similar way.
\begin{equation}
	H_{\rm diag} = H_{\rm diag}^{(K)} + H_{\rm diag}^{N-K}
	\label{eq:H_int}
\end{equation}
where $H_{\rm diag}^{(K)}$ acts on the first $K$ qubits, while $H_{\rm diag}^{N-K}$ is the SM Hamiltonian for $N-K$ qubits. 
The exact expression of $H_{\rm diag}^{(K)}$ is written in the Supplementary Material. 
Due to the structure of the $H_{\rm diag}$ term \eqref{eq:H_int}, the diagonal parts of the effective Hamiltonian \eqref{eq:Matrix_H} are constant plus energy for $N-K$ system size:
\begin{equation}
    \begin{gathered}
	\bra{\Psi_{N-K}}\bra{K}H_{\rm diag}\ket{K}\ket{\Psi_{N-K}} = 
	\\
	=\bra{K}H_{\rm diag}^{(K)}\ket{K} + \bra{\Psi_{N-K}}H_{\rm diag}^{N-K}\ket{\Psi_{N-K}}=
	\\
	=const + E_{N-K}
    \end{gathered}
    \label{eq:rec}
\end{equation}
where $E_M$ is the ground state energy for size $M$.
This constant does not depend on $N$ in the zero sector for an even number of fixed spins and +1 sector for odd (for details, see Supplementary Information).

Due to the recursive structure of Hamiltonian \eqref{eq:rec}, the fractal basis \eqref{eq:ansatz_4part} generates the following effective Hamiltonian $(\hat{H}^R_N)_{ij} = \braket{\phi_i|\hat{H}|\phi_j}$  where $\phi_i$ are the basis vectors:
\begin{equation}
    \begin{gathered}
    \hat{H}_{N}^R = \\
        \begin{pmatrix}
            E_{N-4} + 3 + 2\mu & x A_N& 0 & 0\\
            x A_N& E_{N-2} +1+2\mu &x &0\\
            0 & x & E_{N-2} &xC_N\\
            0 & 0 & xC_N&E_{N-3} + 1+ 2\mu\\
        \end{pmatrix}
    \end{gathered}
    \label{eq:Matrix_H}
\end{equation}
here we have introduced the overlaps $A_N = \bra{01}\braket{\Psi_{N-4}|\Psi_{N-2}} = W_{01}^{N-2} $, and $C_N = \bra{1}\braket{\hat{T}\Psi_{N-3}|\Psi_{N-2}} = f_{N-2} $. 
\begin{equation*}
    f_{N-2} = 
    W_{10}^{N-2}(W_{10}^{N-3} f_{N-4} + W_{110}^{N-3} W_{01}^{N-4})\\ + W_{110}^{N-2}W_{01}^{N-3}.
\end{equation*}
The nondiagonal elements are defined by the hopping part of the Hamiltonian acting on the first fixed qubits.

 \begin{figure*}[ht]
	\centering
	\includegraphics[width=0.8\linewidth]{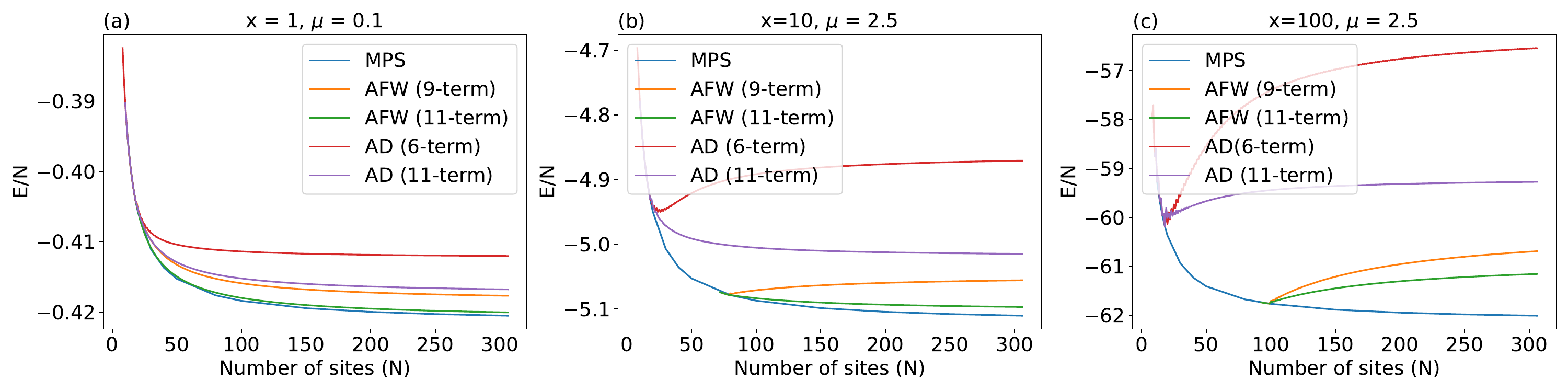}
	\caption{GS energy estimations with the number of sites for different $x$ and $\mu$. Results are obtained using the MPS algorithm, ansatz with fixed weights (AFW) and ansatz diagonalization procedure (AD), both AD and AFW contain either 6,9 or 11 terms.}
	\label{fig:energy}
\end{figure*}
The smallest eigenvalue of matrix~\eqref{eq:Matrix_H} gives the estimation of the GS energy and the corresponding eigenvector provides values of weights $(W_{0011}^N, W_{01}^N, W_{10}^N,W_{110}^N)$ for $N$ qubits. 
Starting from the exact solution for small systems from ED, the ground state energy estimation can be obtained recursively for any system size. 
We refer this method to as Ansatz Diagonalization (AD).
The algorithm for AD is presented in Supplementary Information.

However, in the general case, the weights converge to constants starting from some system size which depends on the parameters $x$ and $\mu$ (see Supplementary Information, Fig.~[4]). 
This system size lies in the range accessible to ED calculations 
only for small $x$ and large $\mu$. When the system approaches to the continuum limit (large $x$) the weight convergence happens much further than in the previous case. 
Thus, the only way to calculate weights at that point is to utilize MPS calculations. 
In any case, we can start from this particular point, and repeat the procedure of calculating energy, assuming that weights are fixed. 
Then the simple formula for the energy is following from $\braket{\Psi_N^A|\hat{H}_N^R|\Psi_N^A}$ where we assume that the values of the weights are the converged ones and does not change with system size. 
We refer this algorithm to as Ansatz with Fixed Weights (AWF). 
The procedure of calculating energy can be improved by including more terms.  We provide more details in the Supplementary Information.

The results of our GS energy estimation are presented in Fig.~\ref{fig:energy}. 
We observe that ansatzes with a larger number of terms 
lead to a better approximation of the GS energy. 
Our fractal algorithms demonstrate the best performance for larger lattice steps (small $x$) and bigger masses $\mu$. 
AFW methods show slightly better energy approximations compared to the AD algorithm. 
However, the starting point of the AFW procedure demands MPS calculations for $x = 10,100$, while the only information needed to launch the AD algorithm is obtained by ED.  

We observe two different energy behaviors in Fig.~\ref{fig:energy}, where some curves decrease monotonically, while others [Fig.~\ref{fig:energy}(b,c)] exhibit non-monotonical behavior.
Cusp points, where the monotonic behavior is broken, correspond to starting points of our algorithm.
We interpret this behavior as indicative of an insufficient ansatz, which needs to include more terms.
The validity of this argument is supported by data illustrated in Fig.~\ref{fig:energy}(b). 
The monotonical pattern of the energy curve appears for the AD 11 term ansatz in contrast to the AD 6 term ansatz. 
The same behavior is observed for AFW 11 and 9 term ansatzes. 
However, the GS energy behavior (Fig.~\ref{fig:energy}(a) 11-term ansatz)  resembles results by MPS for large $N\approx{300}$ number of sites.

Our method provides an upper bound on the energy of the GS, as the procedures are variational. 
There are two related arguments for the deviation between our ansatz and MPS estimations. 
First, different regions of the GS do not perfectly coincide with wave functions for smaller dimensions.
Second, some amplitudes, which are not covered by our ansatz, can contribute to the GS energy. 
However, our estimations do not diverge with the growing system size even in the presence of discrepancies described above.    

{\it Conclusion and outlook}.
Specific properties of quantum many-body systems that potentially could help to overcome the curse of dimensionality challenge in their simulation are of enormous interest.
Here we have expanded on how the fractal property for the ground state of the SM can be used for its analysis.
Specifically, we have used the fractal structure to introduce the representation of the wave function in terms of the system's states for a smaller number of sites.
Despite the simplicity of the approach, we have demonstrated that our fractal anzats give very accurate results for both the ground-state wave function, where we compared our results with the exact diagonalization up to 24 sites,
and the ground state energy. 
In the latter case, we see our predictions are very close (accuracy $99.75\%$) to the results of the state-of-the-art MPS method even for a large ($N\approx{300}$) number of sites in the case of big lattice step ($x=1$). 
We note that the fractal anzats is computationally efficient (all the results are obtained with the use of modest computational resources of a standard PC) and has the feature of controlling its accuracy by increasing the number of terms.

We expect that the fractal anzats can be used to describe a broader class of systems, for example, in condensed matter.
It is interesting to reveal where the fractal property preserves for systems with other types of long-range interactions.
Specifically, it seems to be promising to establish relations to fractal states of quantum matter, such as fractal spin liquids~\cite{Yoshida2013} and Floquet-induced multifractality~\cite{Bandyopadhyay2010,Sinha2018,Sarkar2021,Bagrov2021}.
These cases require however a more detailed study, are subjects of further research.

{\it Acknowledgements}.
We thank A. Bargov and I. Khaymovich for fruitful discussions and for useful comments.
M.C.B. thanks S. K\"uhn for discussions about the phase structure of the model. 
A.K.F. thanks V. Gritsev and A. Garkun for their insightful comments. 
E.V.P., E.S.T., and A.K.F. thank the support of the Russian Science Foundation Grant No. 20-42-05002 (studying the fractal ansatz) and the Russian Roadmap on Quantum Computing (calculating on ground-state energies).
M.C.B. was partly funded by the Deutsche Forschungsgemeinschaft (DFG, German Research Foundation) under Germany's Excellence Strategy -- EXC-2111 -- 390814868.
The code that supports the findings of this study is available on GitHub~\cite{github}.

\bibliography{references.bib}

\end{document}


\begin{center}
\textbf{SUPPLEMENTARY INFORMATION}
\end{center}

\section{Ansatz}
Here we explain in detail how we use fractal structure to obtain distributions and find energies. 
We use several levels of approximation for our calculations.
The first level is given by the following 4-term ansatz:
\begin{equation}
	\begin{gathered}
        \ket{\Psi_N^A} = W_{0011}\ket{0011}\ket{ \Psi_{N-4}} + W_{01}\ket{01}\ket{\Psi_{N-2}}
        +W_{10}\ket{10}\ket{\Psi_{N-2}} + W_{110}\ket{110}\hat{T}\ket{\Psi_{N-3}}.
    \end{gathered}
    \label{eq:ansatz_1}
\end{equation}
where $\hat{T} = \otimes_k \hat{\sigma}^x_k$ is the bit flipping operator.

The second level of approximation can be presented as follows:
\begin{equation}
	\begin{gathered}
        \ket{\Psi_N^A} = 
        W_{001011}\ket{001011}\ket{\Psi_{N-6}} 
        + W_{0011}\ket{0011}\ket{\Psi_{N-4}}+ W_{01}\ket{01}\ket{\Psi_{N-2}} 
        +W_{10}\ket{10}\ket{\Psi_{N-2}} +\\+ W_{110}\ket{110}\hat{T}\ket{\Psi_{N-3}} + W_{11100}\ket{11100}\hat{T}\ket{\Psi_{N-5}}.
    \end{gathered}
    \label{eq:ansatz_2}
\end{equation}
To obtain a more accurate distribution, we add terms with GS for $N-6$ and $N-5$ number of sites.
Then, we use the following 9-term fractal ansatz:
\begin{equation}
	\begin{gathered}
        \ket{\Psi_N^A} = 
        W_{001011}\ket{001011}\ket{\Psi_{N-6}} 
        + W_{0011}\ket{0011}\ket{\Psi_{N-4}}+ W_{01}\ket{01}\ket{\Psi_{N-2}} +\\
        +\ket{10}(W_{100011}\ket{0011}\ket{\Psi_{N-6}} + W_{1001}\ket{01}\ket{\Psi_{N-4}} + W_{1010}\ket{10}\ket{\Psi_{N-4}} + W_{10110}\ket{110}\hat{T}\ket{\Psi_{N-5}}) +\\+ W_{110}\ket{110}\hat{T}\ket{\Psi_{N-3}} + W_{11100}\ket{11100}\hat{T}\ket{\Psi_{N-5}}, 
    \end{gathered}
    \label{eq:ansatz_3}
\end{equation}
$P_{10}\ket{\Psi}$ part is not perfectly fit the wave function for $N-2$ spins, that is why we divide it into four parts.
To include more amplitudes, we propose the 11-term fractal ansatz:
\begin{equation}
	\begin{gathered}
        \ket{\Psi_N^A} = 
        W_{001011}\ket{001011}\ket{\Psi_{N-6}} 
        + W_{0011}\ket{0011}\ket{\Psi_{N-4}}+ W_{01}\ket{01}\ket{\Psi_{N-2}} -\\
        +\ket{10}(W_{10001011}\ket{001011}\ket{\Psi_{N-8}}+ W_{100011}\ket{0011}\ket{\Psi_{N-6}} + W_{1001}\ket{01}\ket{\Psi_{N-4}} +\\+ W_{1010}\ket{10}\ket{\Psi_{N-4}} + W_{10110}\ket{110}\hat{T}\ket{\Psi_{N-5}} + W_{1011100}\ket{11100}\hat{T}\ket{\Psi_{N-7}}) +\\+ W_{110}\ket{110}\hat{T}\ket{\Psi_{N-3}} + W_{11100}\ket{11100}\hat{T}\ket{\Psi_{N-5}}.
    \end{gathered}
    \label{eq:ansatz_4}
\end{equation}
Here more parts in $P_{10}\ket{\Psi}$ sector are added, so the 11-term ansatz includes 9-term.

The following table illustrate the structure and relation of the different ansatzes:
\begin{table*}[ht]
\begin{tabular}{|c|c|c|c|}
\hline
11-term                          & 9-term                         & 6-term                                      & 4-term                                                                                       \\ \hline
$\ket{001011}\ket{\Psi_{N-6}}$   & $\ket{001011}\ket{\Psi_{N-6}}$ & $\ket{001011}\ket{\Psi_{N-6}}$              &                                                                                              \\ \hline
$\ket{0011}\ket{\Psi_{N-4}}$     & $\ket{0011}\ket{\Psi_{N-4}}$   & $\ket{0011}\ket{\Psi_{N-4}}$                & $\ket{0011}\ket{\Psi_{N-4}}$                                                                 \\ \hline
$\ket{01}\ket{\Psi_{N-2}}$       & $\ket{01}\ket{\Psi_{N-2}}$     & $\ket{01}\ket{\Psi_{N-2}}$                  & $\ket{01}\ket{\Psi_{N-2}}$                                                                   \\ \hline
$\ket{10001011}\ket{\Psi_{N-8}}$ &                                & \multirow{6}{*}{$\ket{10}\ket{\Psi_{N-2}}$} & \multirow{6}{*}{\begin{tabular}[c]{@{}c@{}}$\ket{10}\ket{\Psi_{N-2}}$\end{tabular}} \\ \cline{1-2}
$\ket{100011}\ket{\Psi_{N-6}}$   & $\ket{100011}\ket{\Psi_{N-6}}$ &                                             &                                                                                              \\ \cline{1-2}
$\ket{1001}\ket{\Psi_{N-4}}$     & $\ket{1001}\ket{\Psi_{N-4}}$   &                                             &                                                                                              \\ \cline{1-2}
$\ket{1010}\ket{\Psi_{N-4}}$     & $\ket{1010}\ket{\Psi_{N-4}}$   &                                             &                                                                                              \\ \cline{1-2}
$\ket{10110}\hat{T}\ket{\Psi_{N-5}}$   & $\ket{10110}T\ket{\Psi_{N-5}}$ &                                             &                                                                                              \\ \cline{1-2}
$\ket{1011100}\hat{T}\ket{\Psi_{N-7}}$ &                                &                                             &                                                                                              \\ \hline
$\ket{110}\hat{T}\ket{\Psi_{N-3}}$     & $\ket{110}\hat{T}\ket{\Psi_{N-3}}$   & $\ket{110}\hat{T}\ket{\Psi_{N-3}}$                & $\ket{110}\hat{T}\ket{\Psi_{N-3}}$                                                                 \\ \hline
$\ket{11100}\hat{T}\ket{\Psi_{N-5}}$   & $\ket{11100}\hat{T}\ket{\Psi_{N-5}}$ & $\ket{11100}\hat{T}\ket{\Psi_{N-5}}$              &                                                                                              \\ \hline
\end{tabular}
\caption{Each column consists of basis elements, which define ansatzes with 4, 6, 9, and 11 terms, correspondingly. The bit string defines a fixed state for the first few qubits 
from the beginning, $\ket{\Psi_M}$ is the GS for $M$ qubits.}
\label{table:ansatz}
\end{table*}

\section{Recursive Hamiltonians}
As a simple example, let us consider the case where we split the Hamiltonian on the part which acts on the first two spins and the rest. 
\begin{equation}
    \begin{gathered}
        H_{\rm diag} = \frac{\mu}{2}\sum_{n = 0}^{N-1}(1 + (-1)^n\sigma^z_n) +\sum_{n = 0}^{N-2}\left(\frac{1}{2}\sum_{l=0}^n(\sigma_l^z + (-1)^l)\right)^2 = 
        \\
        = \frac{\mu}{2}\sum_{n = 0}^{N-1}(1 + (-1)^n\sigma^z_n)+\frac{1}{4}\sum_{n=0}^{N-2}\sum_{j,j' = 0}^{n}(\sigma^z_{j} + (-1)^j)(\sigma_{j'}^z+(-1)^{j'}) =
        \frac{\mu}{2}(1+\sigma^z_0)+\frac{\mu}{2}(1-\sigma^z_1) +
        \\
        +\frac{N-1}{4}(\sigma_0^z +1)^2 + \frac{N-2}{4}(\sigma_1^z -1)^2 + \frac{1}{2}(N-2)(\sigma_0^z +1)(\sigma_1^z -1) + 
        \frac{1}{2}\left((\sigma_0^z +1)+(\sigma_1^z -1)\right)\sum_{n=2}^{N-2}\sum_{j=2}^{n}(\sigma_j^z + (-1)^j) +
        \\
        + \frac{\mu}{2}\sum_{n = 2}^{N-1}(1 + (-1)^n\sigma^z_n)+\frac{1}{4}\sum_{n=2}^{N-2}\sum_{j,j' = 2}^n (\sigma^z_{j} + (-1)^j)(\sigma_{j'}^z+(-1)^{j'})
    \end{gathered}
\end{equation}
The part which acts on the first two spins is colored in red. 
Red part of the Hamiltonian acts on $\ket{01}$, $\ket{10}$ terms of~\eqref{eq:ansatz_1} ansatz. Then the diagonal terms in the effective Hamiltonian
\begin{equation}
    \begin{gathered}
        \bra{\Psi_{N-2}}\bra{01} H_{\rm diag}\ket{01}\ket{\Psi_{N-2}} = E_{N-2} + 1 + 2\mu \\
        \bra{\Psi_{N-2}}\bra{10} H_{\rm diag}\ket{10}\ket{\Psi_{N-2}} = E_{N-2}
    \end{gathered}
\end{equation}
For any number of fixed spins, the diagonal part of Hamiltonian with zero mass can be rewritten in the following form:
\begin{equation}
    \begin{gathered}
        \sum_{i=0}^{K-1}\frac{N-i}{4}(\sigma_i^z +(-1)^i)^2 + \sum_{i<j =0}^{K-1}\frac{N-1-j}{2}(\sigma_i^z+(-1)^i)(\sigma_j^z+(-1)^j) + \\+\frac{1}{2}\sum_{i=0}^{K-1}(\sigma_i^z+(-1)^i)\sum_{n=K}^{N-2}\sum_{j=K}^{n}(\sigma_j^z+(-1)^j) + \frac{1}{4}\sum_{n=K}^{N-2}\sum_{j,j'=K}^{n}(\sigma_j^z+(-1)^j)(\sigma_{j'}^z+(-1)^j) \equiv \\
        \equiv H_1 + H_2 + H_3 + H_4
    \end{gathered}
    \label{eq:firstK}
\end{equation}
Here, the nonlocal term of the Hamiltonian is divided into four parts. Since we are working in the zero sector for even K or sector 1/2 for the odd number of fixed spins, the following relations are valid
\begin{equation}
    \begin{gathered}
         \bra{K}H_1 + H_2 \ket{K} = {\rm const} \neq f(N)\\
        \bra{K} H_3 \ket{K} = 0\\
    \end{gathered}
\end{equation}
where the $\ket{K}$ denotes the spin configuration of the length $K$ from the corresponding sectors. The constant coming from these expressions does not depend on the system size $N$. The term $H_4$ represents the diagonal part of the Hamiltonian for $N-K$ spins.
\subsection{AD}
In the main part we used 2 methods to find the GS energy, here is the algorithm for the AD method:
\begin{itemize}
	\item \textbf{Step 0.} Find initial weights, overlaps, and GS energies using ED up to $N_0$ number of spins.
	\item \textbf{Step 1.} Construct the reduced Hamiltonian matrix for $N_0+1$ spins in the basis generated by the fractal ansatz.
	\item \textbf{Step 2.} Find the minimal eigenvalue and corresponding eigenvector of that matrix for $N_0+1$ spins.
	\item \textbf{Step 3.} Save the weights, which are elements of the minimal eigenvector, and the minimal eigenvalue, and calculate overlaps from the weights for $N_0+1$ spins.
	\item \textbf{Step 3.} Repeat from step 1 increasing the initial system size $N_0\rightarrow N_0+1$.
\end{itemize}
For each step of the algorithm above, we calculate weights recursively.

For ansatz diagonalization method we build Hamiltonian in a corresponding basis. 
Nonzero elements of the reduced Hamiltonian for ansatz~\eqref{eq:ansatz_2} are as follows:
\begin{equation}
    \begin{gathered}
            H_{1,1} = E_{N-6} + 5 + 2\mu, \hspace{3cm}
            H_{2,2} = E_{N-4} + 3 + 2\mu, \\
            H_{3,3} = E_{N-2} + 1 + 2\mu, \hspace{3cm}
            H_{4,4} = E_{N-4},\;\;\;\;\;\;\;\;\;\;\;\;\;\;\;\\
            H_{5,5} = E_{N-3}+1+2\mu, \hspace{3cm}
            H_{6,6} = E_{N-5}+3 + 2\mu,\\
            H_{1,2} = \braket{001011\Psi_{N-6}|\hat{H}|0011\Psi_{N-4}} = xW_{01}^{N-4},\\
            H_{1,3} = \braket{001011\Psi_{N-6}|\hat{H}|01\Psi_{N-2} }= xW_{0011}^{N-2},\\
            H_{2,3} = \braket{0011\Psi_{N-4}|\hat{H}|01\Psi_{N-2}} = xW_{0011}^{N-2},\\
            H_{3,4} =  \braket{01\Psi_{N-2}|\hat{H}|10\Psi_{N-2}}= x,\\
            H_{4,5} = \braket{10\Psi_{N-2}|\hat{H}|110\hat{T}\Psi_{N-3}} = x f_{N-2}, \\
            H_{5,6} =\braket{110\hat{T}\Psi_{N-3}|\hat{H}|11100\hat{T}\Psi_{N-5}} =  xW_{01}^{N-3},\\
            f_{N-2} = x(W_{10}^{N-2}(W_{10}^{N-3}f_{N-4}+W_{110}^{N-3}W_{01}^{N-4} + W_{110}^{N-2}W_{11100}^{N-4}) + W_{11100}^{N-2}W_{0011}^{N-3}).
    \end{gathered}
    \label{H:AD}
\end{equation}
The rest elements in the reduced Hamiltonian are zero.

The reduced Hamiltonian for the 11-term ansatz is the following: 
\begin{equation}
    \begin{gathered}
            H_{4,4} = E_{N-8} + 5 + 2\mu,\hspace{3cm}
            H_{5,5} = E_{N-6} + 3 + 2\mu,\\
            H_{6,6} = E_{N-4} + 1+ 2\mu,\hspace{3cm}
            H_{7,7} = E_{N-4},\;\;\;\;\;\;\;\;\;\;\;\;\;\;\;\\
            H_{8,8} = E_{N-5} +1+2,\hspace{3cm}
            H_{9,9} = E_{N-7}+3+2\mu.
    \end{gathered}
\end{equation}
The following elements replace $H_{3,4}$ and $H_{4,5}$:
\begin{equation}
    \begin{gathered}
        H_{3,4} = \braket{01\Psi_{N-2}|H|10001011\Psi_{N-8}} = xW_{001011}^{N-2},\\ 
        H_{3,5} = \braket{01\Psi_{N-2}|H|100011\Psi_{N-6}} = xW_{0011}^{N-2}, \\
        H_{3,6} = \braket{01\Psi_{N-2}|H|10001\Psi_{N-4}} = xW_{01}^{N-2},\\
        H_{3,7} = \braket{01\Psi_{N-2}|H|1010\Psi_{N-8}} = xW_{10}^{N-2} = g_{N-2},\\
        H_{3,8} = \braket{01\Psi_{N-2}|H|10110\Psi_{N-5}} = xW_{110}^{N-2},\\
        H_{3,9} = \braket{01\Psi_{N-2}|H|1011100\Psi_{N-7}} = xW_{11100}^{N-2},\\
        H_{7,10} = \braket{1010\Psi_{N-4}|\hat{H}|110\hat{T}\Psi_{N-3}} = f_{N-3},\\
        H_{8,10} = \braket{10110\hat{T}\Psi_{N-5}|\hat{H}|110\hat{T}\Psi_{N-3}} = xW_{01}^{N-2},\\
        H_{9,10} = \braket{1011100\hat{T}\Psi_{N-7}|\hat{H}|110\hat{T}\Psi_{N-3}} = xW_{0011}^{N-3},\\
        H_{4,5} = \braket{10001011\Psi_{N-8}|H|100011\Psi_{N-6}} = xW_{01}^{N-6},\\
        H_{4,6} = \braket{10001011\Psi_{N-8}|H|1001\Psi_{N-4}} = xW_{0011}^{N-4},\\
        H_{8,9} = \braket{10110\hat{T}\Psi_{N-5}|H|1011100\hat{T}\Psi_{N-7}} = xW_{01}^{N-5}\\
        g_{N}  = \braket{10\Psi_{N-2}|\Psi_N} = W_{10001011}^NW_{001011}^{N-2} + W_{100011}^{N}W_{0011}^{N-2}\\ + W_{1001}^{N}W_{01}^{N-2} + W_{1010}g_{N-2} + W_{10110}^NW_{110}^{N-2} + W_{1011100}^{N}W_{11100}^N,\\
        f_N = W_{100011}^NW_{11100}^{N-1} + W_{1001}^NW_{110}^{N-1} + W_{1010}^Np_{N-1} +\\+ W_{10110}^NW_{1001}^{N-1} + W_{1011100}^NW_{100011}^{N-1} + W_{110}^NW_{01}^{N-1} + W_{11100}^NW_{0011}^{N-1},\\
        p_N = \braket{\Psi_{N}|101\hat{T}\Psi_{N-3}} = W_{1010}^N f_{N-3} + W_{10110}^N W_{01}^{N-3} + W_{1011100}^N W_{0011}^{N-3}.
    \end{gathered}
    \label{H:AD2}
\end{equation}
Elements $H_{1,1}$, $H_{2,2}$, $H_{3,3}$, $H_{1,2}$, $H_{1,3}$, and $H_{2,3}$ are the same as for matrix (\ref{H:AD}). 
Elements $H_{10,11}$, $H_{10,10}$,$H_{11,11}$ of matrix (\ref{H:AD2}) are equal to the elements $H_{5,6}$, $H_{5,5}$, $H_{6,6}$ of matrix (\ref{H:AD}).
\subsection{AWF}

The procedure of finding the GS energy using the ansatz with Fixed Weights (AFW) consists of the following steps.
\begin{itemize}
	\item \textbf{Step 0.} Choose the initial system size $N_0$ by observing the behavior of the weights with the number of spins. At the starting point, weights have to converge. 
	\item \textbf{Step 1.} Calculate the weights, overlaps, and energies up to $N_0$ spins according to the ansatz using exact diagonalization or MPS approach.
	\item \textbf{Step 2.} Calculate energy using formula (\ref{eq:energy}) for $N_0+1$ system size.
	\item \textbf{Step 3.} Save energy for $N_0+1$ spins and repeat from step 2 by reassigning initial system size $N_0\rightarrow N_0+1$.
\end{itemize}
For the AWF method we use the following formulas:
\begin{equation}
    \begin{gathered}
        E_{N} = (W_{0011})^2(E_{N-4} + 3+ 2\mu) +(W_{01})^2(E_{N-2} + 1 +2\mu) + (W_{10})^2E_{N-2} +\\+ (W_{110})^2(E_{N-3} + 1 + 2\mu)+ +2W_{0011}W_{01}\bra{01}\braket{\Psi_{N-4}|\Psi_{N-2}}+ 2W_{01}W_{10}\braket{\Psi_{N-2}|\Psi_{N-2}}+ \\ +2W_{10}W_{110}\bra{1}\braket{\hat{T}\Psi_{N-3}|\Psi_{N-2}}
    \end{gathered}
    \label{eq:energy}
\end{equation}
where $E_{N}$ is the mean value of the Hamiltonian averaged over fractal ansatz (2) in the main text for $N$ qubits. 
Here we assume that the values of the weights are the converged ones, $W_{s}:=\lim_{N\to\infty} W_{s}^{N}$. 
The state preserves its self-similar structure since the weights converge with the system size. We can deduce from this fact that overlaps also become constants along with weights. 

The mean energy in terms of ansatz~(\ref{eq:ansatz_3}) that is used in the AWF method reads
\begin{equation}
	\begin{gathered}
        E_N = 
        W_{001011}^2(E_{N-6} + 5 + 2\mu )
        + W_{0011}^2(E_{N-4} + 3 + 2\mu) + W_{01}^2(E_{N-2} + 1 + 2\mu) +\\
        (W_{100011}^2(E_{N-6}+ 3 +2\mu) + W_{1001}^2(E_{N-4}+1+2\mu) + W_{1010}^2(E_{N-4}) + W_{10110}^2(E_{N-5}+1+2\mu ))+\\  W_{110}^2(E_{N-3}+1+2\mu) + W_{11000}^2(E_{N-5}+3 + 2\mu)  + 2C.
    \end{gathered}
    \label{eq:Energy}
\end{equation}
For ansatz (\ref{eq:ansatz_4}) the energy takes the following form:
\begin{equation}
	\begin{gathered}
        E_N = 
        W_{001011}^2(E_{N-6} + 5 + 2\mu )
        + W_{0011}^2(E_{N-4} + 3 + 2\mu) + W_{01}^2(E_{N-2} + 1 + 2\mu) +\\
        (W_{10001011}^2(E_{N-8} + 5+2\mu) + W_{100011}^2(E_{N-6}+ 3 +2\mu) + W_{1001}^2(E_{N-4}+1+2\mu) + W_{1010}^2(E_{N-4}) +\\
        + W_{10110}^2(E_{N-5}+1+2\mu )+ W_{1011100}^2(E_{N-7} + 3+2\mu) ) + W_{110}^2(E_{N-3}+1+2\mu) + W_{11000}^2(E_{N-5}+3 + 2\mu)  + 2C'.
    \end{gathered}
    \label{eq:Energy}
\end{equation}
Where $C$ comes from the average of the hopping term and has the following form:
\begin{equation}
	\begin{gathered}
        C = W_{001011}W_{01}\braket{001011\Psi_{N-6}|\hat{H}|01\Psi_{N-2} } + 
        W_{001011}W_{0011}\braket{001011\Psi_{N-6}|\hat{H}|0011\Psi_{N-4}}+\\+
        W_{0011}W_{01} \braket{0011 \Psi_{N-4} | \hat{H} | 01 \Psi_{N-2}}
        + W_{01}W_{100011}\braket{01 \Psi_{N-2} | \hat{H} | 100011 \Psi_{N-6}} +\\+
        W_{01}W_{1001} \braket{01 \Psi_{N-2} | \hat{H} | 1001 \Psi_{N-4}} + 
        W_{01}W_{1010}\braket{01 \Psi_{N-2} | \hat{H} | 1010 \Psi_{N-4}} + \\+
        W_{01}W_{10110}\braket{01 \Psi_{N-2} | \hat{H} | 10110 \hat{T} \Psi_{N-5}}+
        W_{1001}W_{100011}\braket{100011 \Psi_{N-6} | \hat{H} | 1001 \Psi_{N-4}}\\+
        W_{1010}W_{10110}\braket{1010 \Psi_{N-4} | \hat{H} | 10110 \hat{T}\Psi_{N-5}}+
        W_{110}W_{1010}\braket{110 \hat{T} \Psi_{N-3} | \hat{H} | 1010 \Psi_{N-4}} +\\+
        W_{110}W_{10110}\braket{110 \hat{T} \Psi_{N-3} | \hat{H} | 10110 \hat{T} \Psi_{N-5}}+
        W_{001011}W_{0011}\braket{001011 \Psi_{N-6} | \hat{H} | 0011 \Psi_{N-4}}+\\+
        W_{001011}W_{01}\braket{001011 \Psi_{N-6} |\hat{H} | 01 \Psi_{N-2}} +
        W_{11100}W_{110}\braket{11100 \hat{T} \Psi_{N-5} | \hat{H} | 110 \hat{T} \Psi_{N-3}} +\\
        +W_{1001}W_{1010}\braket{1001\Psi_{N-4}|1010\Psi_{N-4}}.
    \end{gathered}
    \label{eq:const}
\end{equation}

Eq.~(\ref{eq:const}) can be further simplified, knowing that $\hat{H}$ acts on these states by flipping only neighboring spins. 
Thus, $C$ can be written down in terms of overlaps
\begin{equation}
	\begin{gathered}
	C/x = 
	    W_{001011}W_{01}\braket{0011\Psi_{N-6}|\Psi_{N-2} } + 
        W_{001011}W_{0011}\braket{01\Psi_{N-6}|\Psi_{N-4}}+\\+
        W_{0011}W_{01} \braket{ \Psi_{N-2} | 01\Psi_{N-4} }
        + W_{01}W_{100011}\braket{ \Psi_{N-2} | 0011 \Psi_{N-6}}  +\\+
        W_{01}W_{1001} \braket{ \Psi_{N-2} | 01 \Psi_{N-4} }  + 
        W_{01}W_{1010}\braket{\Psi_{N-2} | 10 \Psi_{N-4} }+ \\+
        W_{01}W_{10110}\braket{ \Psi_{N-2} | 110 \hat{T}\Psi_{N-5}} +
        W_{1001}W_{100011}\braket{01 \Psi_{N-6}|  \Psi_{N-4}} \\+
        W_{1010}W_{10110}\braket{ \Psi_{N-4} | 1 \hat{T}\Psi_{N-5}}+
        W_{110}W_{1010} \braket{ \hat{T}\Psi_{N-3} | 0 \Psi_{N-4} } +\\+
        W_{110}W_{10110}\braket{\hat{T}\Psi_{N-3} | 10 \hat{T}\Psi_{N-5}} +
        W_{001011}W_{0011} \braket{01 \Psi_{N-6}|  \Psi_{N-4}} +\\+
        W_{001011}W_{01}\braket{\Psi_{N-2} | 0011 \Psi_{N-6}}  +
        W_{11100}W_{110}\braket{ \hat{T} \Psi_{N-3} | 10 \hat{T}\Psi_{N-5} }+\\
        +W_{1001}W_{1010}.
    \end{gathered}
    \label{eq:const2}
\end{equation}
\begin{equation}
\begin{gathered}
        C'/x = C/x + W_{10001011}W_{100011}\braket{01\Psi_{N-8}|\Psi_{N-6}} + W_{10001011}W_{1001}\braket{0011\Psi_{N-8}|\Psi_{N-4}}+\\+
        W_{10110}W_{1011100}\braket{\hat{T}\Psi_{N-5}|10\hat{T}\Psi_{N-7}}+ W_{10}W_{10001011}\braket{\Psi_{N-2}|001011\Psi_{N-8}}+\\+
        W_{01}W_{1011100}\braket{\Psi_{N-2}|11100\hat{T}\Psi_{N-7}}+ 
        W_{1011100}W_{110}\braket{1100\hat{T}\Psi_{N-7}|\hat{T}\Psi_{N-3}}.
\end{gathered}
\end{equation}
Using this formula, we can recursively calculate the energy for arbitrary numbers of qubits. 

\section{Qubism framework and fractal image compression}

\begin{figure}[ht]
	\centering
	\includegraphics[scale=0.2]{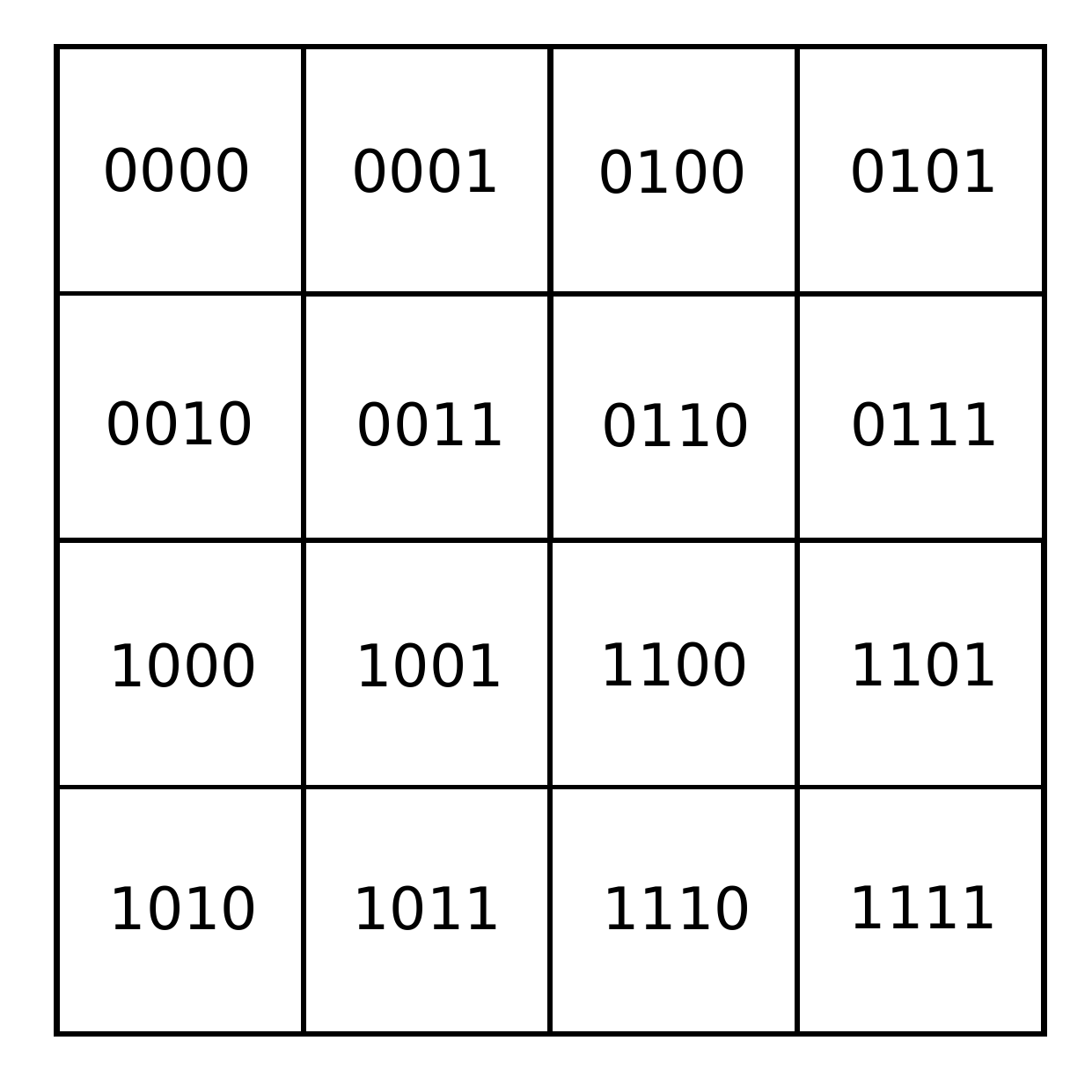}
	\vskip -3mm
	\caption{Qubism representation for a 4 qubit state, in which bitstrings denote qubit configurations of a state. Each rectangle represents a pixel, which value is the corresponding probability to the power of $0.2$.}
	\label{fig:qubism_scheme}
\end{figure}

\begin{figure*}
\centering
\includegraphics[width=1\linewidth]{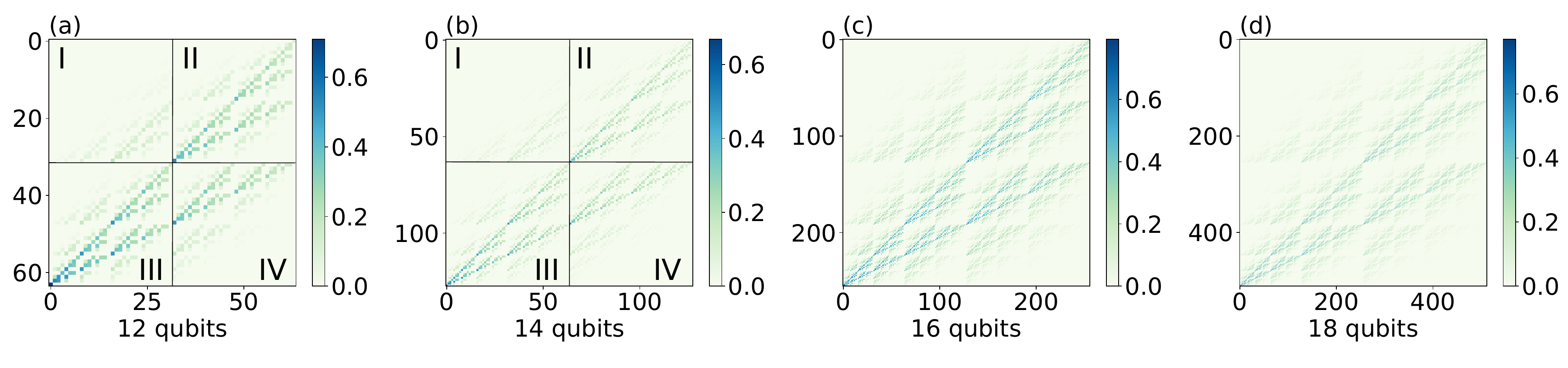}
\vskip-3mm
\caption{Fractal structure of the GS wave functions of the model in the qubism representation for (a) $N=12$, (b) $N=14$, (c) $N=16$, and (d) $N=18$ qubits and parameters $x = 1$, $\mu = 0.1$. 
Regions II and III for 12 and 14 qubit distributions coincide with regions in Fig.~[1], and I and IV regions extend that of Fig.~[1].
The qubism shows the main property: an increase in the image resolution reflects an increase in self-replicating structures.
Here (a,b) are the squared amplitudes to the power of 0.2, while (c,d) is the square amplitudes to the power of 0.1 for a brighter image.
	}
\label{fig:qubism}
\end{figure*}

The qubism framework~\cite{Lewenstein2012}, which establishes the specific mapping between many-body quantum states and 2D plots, provides an alternative way to capture the self-similarity of the model. 
For an even number of sites, qubism maps the wave function onto a square image of size $2^{N/2}\times 2^{N/2}$. 
Each pixel of the image represents the absolute square of the amplitude for a single basis state.
The order of the pixels in the image is such that each row corresponds to one of the $2^{N/2}$ configurations of the spins on even positions, while each column corresponds to a given configuration of the odd spins. 
For example, a four qubit state corresponds to a 4 by 4 image in the qubism representation, see Fig.~\ref{fig:qubism_scheme}. 
We illustrate the fractal structure of the GS wave functions for different values of the system size in Fig.~\ref{fig:qubism}.

Since the qubism framework provides a way to look at a state as an image, one can use a large variety of computer graphics software for obtaining a compact image representation.  
Having this representation, one can decompress the image into a grey-scale image of arbitrary size, thus generating a distribution for a larger system. 
The particular class of such algorithms is fractal-image-compression~\cite{Agarwal2019}. 
The idea is to find transformations that scale larger parts into smaller parts (compression) and vice versa (decompression), such that the parts after the transformation are similar to the corresponding piece of the original image.
Specifically, we use the simplest fractal-compression-decompression package~\cite{pvigier_fractal}, which implements segmentations, compression, and decompression of images.
Post-processing is also required. 
After the decompression step, we apply a mask that nullifies all the amplitudes that violate the total spin conservation and then normalizes the whole state.

\section{Fractal structure in different phases}

\begin{figure*}[ht]
	\centering
    \includegraphics[width=1\linewidth]{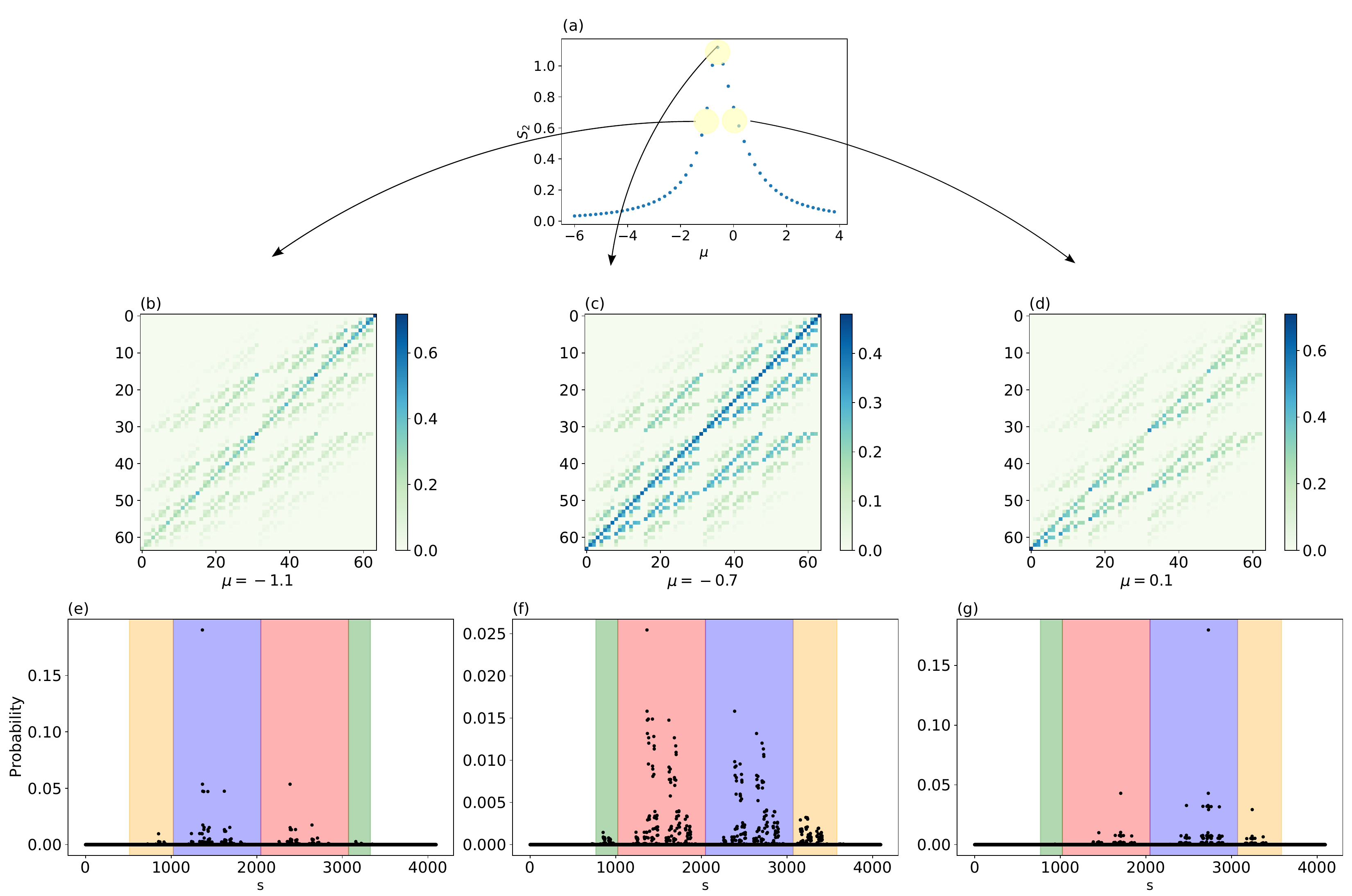}
	\vskip -3mm
	\caption{The phase transition in the system of $N=12$ qubits at $x = 1$ and different $\mu$ is illustrated in the qubism framework.}
	\label{fig:phasetransition}
\end{figure*}

At a certain value of the background field (corresponding to $\epsilon_0=1/2$ in the lattice formulation) and at the critical mass $(m/g)_{\mathrm{c}}\approx 0.33$, 
the SM exhibits a second-order phase transition belonging to the Ising universality class~\cite{Coleman1976,Byrnes2002}.
In the zero charge sector, it is possible to unitarily transform the Hamiltonian such that $\epsilon_0\to \epsilon_0-1/2$ and $\mu\to-\mu$, so that the 
same critical behavior can be observed at zero background field for a negative mass~\cite{Funcke2020}.
This phenomenon has in fact been the object of recent variational quantum simulation experiments~\cite{Kokail2019,Straupe2021}.

In terms of the spin representation, we can understand the different nature of both phases by comparing the ground states in the limits $\mu\to\pm\infty$. 
For $\mu \gg 0$, the energetically favoured configuration is $\ket{0101\ldots 01}$, corresponding to a full filling of fermions and antifermions, 
while for $\mu \ll 0$, the ground state is empty, which corresponds to $\ket{1010\ldots10}$. 

To detect the transition, we can use the R\'enyi entropy density~\cite{Kokail2019,Blatt2019}, which can be defined as follows:
\begin{equation}
	S_2 = -\frac{\log(\sum_i P_i^2)}{N}
	\label{eq:Rentropy}
\end{equation}
where $P_i$ are probabilities of each basis state in the wave function. 
At $x=1$, the value of $\mu$ corresponding to the (negative) critical mass is $\mu \sim -0.7$. 
We show in Fig.~\ref{fig:phasetransition}(a) the value of the R\'enyi entropy density as a function of $\mu$ 
for a system size $N=12$.

The fractal structure of the wave function also essentially changes across the critical point. 
The structure of the qubism representation is common for both phases but reflected around diagonal and antidiagonal in the qubism representation Fig.~\ref{fig:phasetransition}.
These two reflections correspond to the exchanging 0 and 1 in spin configurations. The top right pixel is the brightest one in the phase where $\mu < -0.7$. 
This pixel corresponds to the spin configuration $\ket{0101..01}$ which is the vacuum of 1D QED model. The situation in the other phase is the opposite. 
The brightest one is the bottom left pixel, which is the full filling of electrons on a lattice ($\ket{1010..10}$ spin configuration).
The same behavior is observed for the probability distributions across different phases [see Figs. \ref{fig:phasetransition}(e)--(g)]. 
Our ansatz (2) in the main text properly catches the structure by dividing distributions into four regions. 
This statement is supported visually in Figs. \ref{fig:phasetransition}(e)--(g). 
Thus, the fractal ansatz can be applied to describing GS in different regimes, i.e. far away and close to the region where the phase transition occurs.

\section{Weights}

\begin{figure}[H]
	\centering
	\includegraphics[width=1\linewidth]{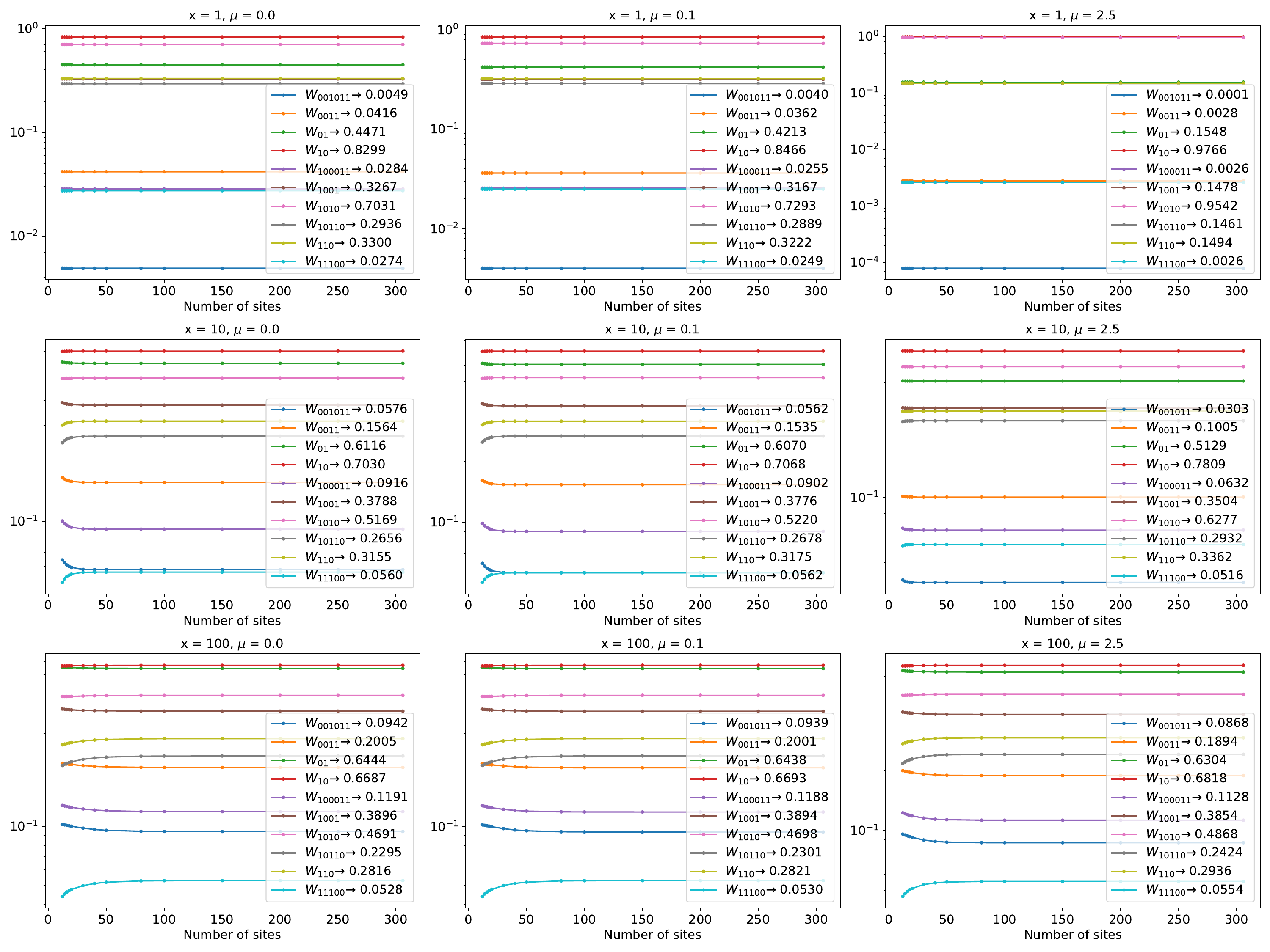}
	\vskip-3mm
	\caption{MPS calculations of weights for different parts of the wave function and various parameters of the Hamiltonian.}
	\label{fig:WeightsMPS}
\end{figure}

Here we present the absolute values of weights for different parameters $x$ and $\mu$ and number of sites obtained by MPS calculations, see Fig.~\ref{fig:WeightsMPS}. 
Weights are required in our approach for calculating wave functions and energies. 
One can see that weights become constants with the system size. 
This convergence is essential for the validity of the self-similar approach.

\section{Errors}

\begin{figure}[H]
	\centering
	\includegraphics[width=1\linewidth]{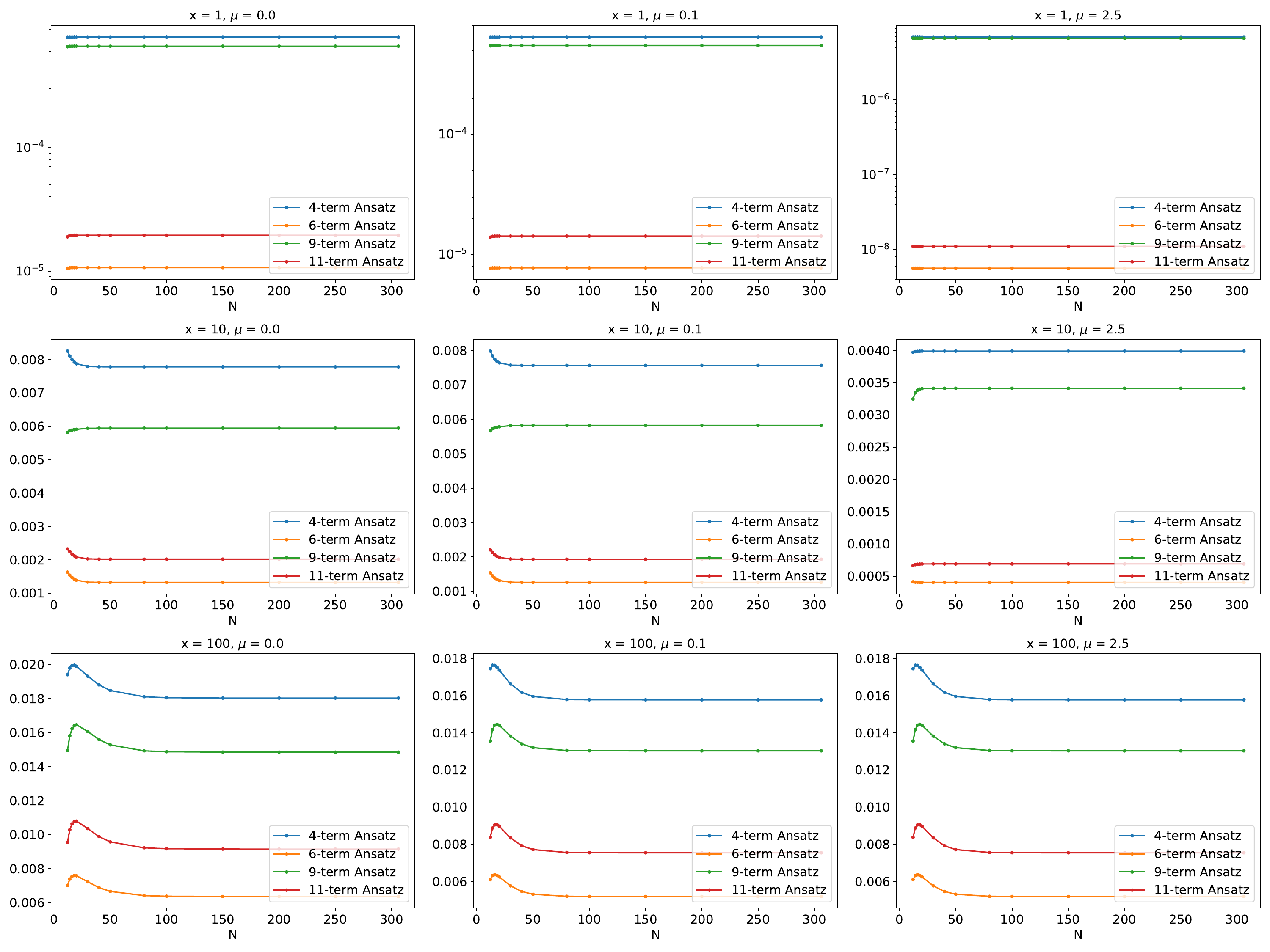}
	\vskip-3mm
	\caption{The figure represents neglected probabilities for different parameters and different ansatzes. Neglected probabilities are equal to the $1-\sum_i{W_i^2}$  where $W_i$ is a weight.}
	\label{fig:Errors}
\end{figure}

The errors that occurred in the algorithm come in two ways. First of all, we neglect some amplitudes using fractal ansatzes. 
Secondly, we simplify the structure of the wave function assuming it to be made of exactly the distributions for previous dimensions. 
The first type of error can be easily calculated. 
In Fig.~\ref{fig:Errors} we illustrate the amount of neglected probability for different parameters and different ansatzes.

\section{Different bases}

\begin{figure}[H]
	\centering
    \includegraphics[width=1\linewidth]{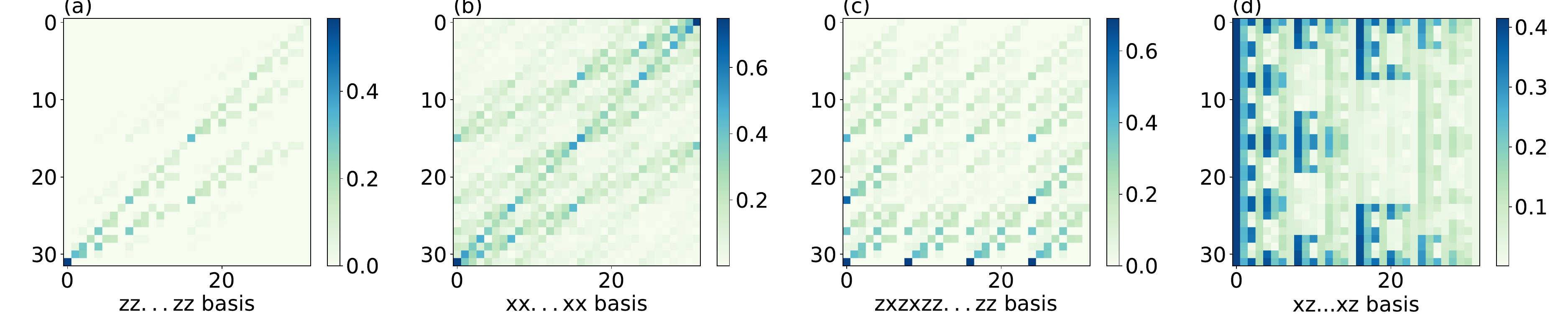}
	\vskip-3mm
	\caption{Probability distributions in different bases, clearly indicate the presence of the fractal structure.}
	\label{fig:basis}
\end{figure}

Here we illustrate the existence of the fractal structure in different bases. 
We note that the fractal structure is unique on each basis. 
It can be easily seen in qubism representation, see Fig.~\ref{fig:basis}.

\bibliographystyle{apsrev4-1}
\bibliography{references-2.bib}